\documentclass[a4paper,twocolumn,english,conference]{ieeeconfcdc}
\usepackage[T1]{fontenc}
\usepackage{babel}
\usepackage{amsmath}
\usepackage{stackrel}
\usepackage{graphicx}
\usepackage[unicode=true,
 bookmarks=true,bookmarksnumbered=true,bookmarksopen=true,bookmarksopenlevel=1,
 breaklinks=false,pdfborder={0 0 0},pdfborderstyle={},backref=false,colorlinks=false]
 {hyperref}
\hypersetup{pdftitle={Your Title},
 pdfauthor={Your Name},
 pdfpagelayout=OneColumn, pdfnewwindow=true, pdfstartview=XYZ, plainpages=false}

\makeatletter

\IEEEoverridecommandlockouts                              % This command is only
\UseRawInputEncoding

\usepackage{cite}

\title{\LARGE \bf
Resilient distributed integral control for multimachine power systems with inherent input constraint satisfaction
}

\author{Theodoros E. Kavvathas, George C. Konstantopoulos, and Charalambos Konstantinou% <-this % stops a space
\thanks{This work was supported under Grant 81359 from the Research Committee of the University of Patras via ''C.CARATHEODORY'' program.
T. E. Kavvathas and G. C. Konstantopoulos are with the Department of Electrical and Computer Engineering, University of Patras, Rion 26504, Greece. C. Konstantinou is with the CEMSE Division, King Abdullah University
of Science and Technology (KAUST), Thuwal 23955-6900, Saudi Arabia.
        {\tt\small  Emails: up1066602@upnet.gr, g.konstantopoulos@ece.upatras.gr, charalambos.konstantinou@kaust.edu.sa}}%
}

\@ifundefined{showcaptionsetup}{}{%
 \PassOptionsToPackage{caption=false}{subfig}}
\usepackage{subfig}
\makeatother

\begin{document}
\maketitle\thispagestyle{empty} \pagestyle{empty}

%%%%%%%%%%%%%%%%%%%%%%%%%%%%%%%%%%%%%%%%%%%%%%%%%%%%%%%%%%%%%%%%%%%%%%%%%%%%%%%%

\begin{abstract}
In this paper, a novel distributed controller for multimachine power
systems is proposed to guarantee grid frequency restoration and accurate
real and reactive power sharing among the generator units, while maintaining
the generator inputs (mechanical torque and field excitation voltage)
within given bounds. The boundedness of the controller outputs (generator
inputs) is rigorously proven using vector field theory. It is additionally
shown that even if one generator input reaches its upper/lower limit,
the remaining units can still accomplish the desired control tasks
without modifying the controller structure or dynamics; hence introducing
enhanced system resilience using the proposed approach. This has been
accomplished in a unified control structure while using neighbour-to-neighbour
communication, thus maintaining the distributed nature of the controller.
An example of a 10-bus, 4-machine power system is simulated to verify
the proposed controller performance under sudden changes of the load
demand.
\end{abstract}
%%%%%%%%%%%%%%%%%%%%%%%%%%%%%%%%%%%%%%%%%%%%%%%%%%%%%%%%%%%%%%%%%%%%%%%%%%%%%%%%

\section{Introduction}

Modern electricity networks integrate monitoring, control, and communication
technologies into the physical infrastructure (power network, i.e.,
energy resources, loads, transmission/distribution lines, etc.); thus
forming a cyber-physical architecture \cite{en13092169,7122365,6011696}.
The bidirectional flow of power and information among the energy resources
(conventional generators, distributed energy resources - DERs, flexible/responsive
loads) introduces advanced control and optimisation properties for
the power system that can be utilised in a centralised, decentralized,
or distributed manner \cite{6423237}.

Although the integration of DERs (renewable energy systems, energy
storage systems, intelligent loads) continuously increases driven
by efforts of grid decarbonisation strategies in different regions
of the world, at the transmission level, conventional synchronous
generators still play a key role in maintaining system stability (frequency
and voltage regulation) \cite{898110,9127527}. Hence, a multimachine
power network should be suitably coordinated to ensure restoration
of the system frequency at the rated value and optimal load flow,
which is often achieved by proportionally sharing real and reactive
power among the generators \cite{6423237,8943305}.

Since centralised coordination of multiple synchronous generators
becomes more and more vulnerable to cyber attacks \cite{ge2022resilience},
distributed control design based on neighbour-to-neighbour communication
has been widely established as a suitable alternative \cite{9420296}.
In such frameworks, the design of resilient networked control has
been extensively studied under scenarios of loss of communication,
package drop, and time varying delays \cite{9420296,6194234,8848621,8966514}.
Additional scenarios that can lead the entire multimachine power system
to instability include sudden changes of the load demand \cite{amini2016dynamic},
sensor/actuator faults \cite{8935105}, or setpoint attacks \cite{9993031}.
Although several resilient controllers have been introduced in literature,
in the majority of the cases, the physical/technical/operational constraints
of the synchronous generators, e.g., actuator limits (mechanical torque
or field excitation voltage limits), are not taken into consideration.
These constraints can be handled by typical saturation units, but
as it has been shown in \cite{9837149,4596530}, conventional distributed
integral control with saturation units at the mechanical torque input
of each generator can lead to instabilities. Hence, according to the
authors' knowledge, a resilient distributed control approach that
takes into consideration all of the input constraints of each generator
and continues to regulate the frequency at its nominal value and maintains
the power sharing, even when at least one of the generator is forced
to reach its operating limits, has not been developed yet.

In this paper, a novel distributed control design for multimachine
power networks is proposed that has a unique and unified structure
under both normal operation and under scenarios that cause some of
the generators to operate at their limits (e.g., sudden change of
load demand). Using vector field analysis, it is analytically proven
that the proposed distributed controller maintains the controller
outputs (generator mechanical torque and field excitation voltage
inputs) within a given range. At the same time, the controller ensures
grid frequency restoration to the rated value and accurate real and
reactive power sharing among the generator units based only on neighbour-to-neighbour
communication. This unique control scheme is resilient in the sense
that the desired tasks (frequency restoration and power sharing) are
still accomplished among the generators which operate within their
bounded range, even if one or more of the generators force their inputs
to the upper or lower limit. As demonstrated in this work, the controller
does not need to modify its structure to accomplish it; hence offering
a unified solution under normal and abnormal conditions. The effectiveness
of the proposed control is verified using a 10-bus, 4-machine power
network under sudden changes of the load demand real and reactive
power.

The paper is structured as follows: In Section \ref{sec:Problem-formulation},
the dynamic model of the multimachine power system is presented and
the main control problem is formulated. In Section \ref{sec:Proposed-control-architecture},
the proposed structure of the resilient distributed controller is
presented and mathematically analysed. The boundedness of the controller
states is rigorously proven and it is analytically shown how the controller
accomplishes accurate frequency restoration to the rated value and
both real and reactive power sharing among the generators, while ensuring
that the generator inputs remain within some predefined ranges at
all times. Simulation results of the proposed scheme are given in
Section \ref{sec:4Simulation-results} and the final conclusions are
provided in Section \ref{sec:5Conclusions}.

\section{Problem formulation\label{sec:Problem-formulation}}

\subsection{System modelling\label{subsec:System-modelling}}

We consider a generic multimachine power network with $N$ buses and
$n$ generators where, for simplicity, the $i-$th generator is described
by the well-known 3rd-order dynamic model:
\begin{align}
\frac{d\delta_{i}}{dt} & =\omega_{b}(\omega_{i}-\omega_{s})\label{eq:delta}\\
2H_{i}\frac{d\omega_{i}}{dt} & =T_{mi}-T_{ei}-D_{i}(\omega_{i}-\omega_{s})\label{eq:omega}\\
T'_{d0i}\frac{dE'_{qi}}{dt} & =E_{fi}-E'_{qi}+(X_{di}-X'_{di})i_{di}\label{eq:Eq}
\end{align}
where $\delta_{i}$ is the rotor angle, $\omega_{b}$ is the base
angular velocity, $\omega_{i}$ is the generator angular velocity,
$\omega_{s}$ is the synchronous angular velocity (i.e., $\omega_{i}-\omega_{s}$
defines the slip), $H_{i}$ is the inertia constant, $D_{i}$ is the
generator motor damping, $T_{mi}$ and $T_{ei}$ are the mechanical
and electrical torques, respectively, $T'_{d0i}$ is the $d-$axis
open circuit transient time constant, $E'_{qi}$ is the $q-$axis
transient emf, $E_{fi}$ is the field excitation voltage, $X_{di}$
and $X'_{di}$ are the $d-$axis synchronous and transient reactances,
and $i_{di}$ is the $d-$axis component of the stator current. Note
that the electrical torque is given as 
\begin{equation}
T_{ei}=\psi_{di}i_{qi}-\psi_{qi}i_{di},\label{eq:Te}
\end{equation}
where $i_{qi}$ is the $q-$axis component of the stator current,
$\psi_{di}$ and $\psi_{qi}$ are the $d-$ and $q-$axis flux of
the damper windings, respectively, and are linked to the stator currents
and the $q-$axis transient emf as
\begin{align}
\psi_{di} & =X'_{di}i_{di}+E'_{qi}\label{eq:psid}\\
\psi_{qi} & =X'_{qi}i_{qi},\label{eq:psiq}
\end{align}
where $X'_{qi}$ is the $q-$axis transient reactance. Furthermore,
the relationship between the generator stator current and voltages
is
\begin{align}
i_{qi} & =\frac{1}{r_{ai}}(\psi_{di}-v_{qi})\label{eq:iq_vq}\\
i_{di} & =\frac{1}{r_{ai}}(-\psi_{qi}-v_{di})\label{eq:id_vd}
\end{align}
where $r_{ai}$ is the armature resistance.

In order to obtain the model of the entire multimachine power network,
we need to rotate the generator current and voltages by the rotor
phase angle $\delta_{i}$, resulting in the generator current and
voltages in a common reference frame, i.e.,

\begin{align}
i_{Qi}+ji_{Di} & =(i_{qi}+ji_{di})e^{j\delta_{i}}\label{eq:currents_DQ}\\
v_{Qi}+jv_{Di} & =(v_{qi}+jv_{di})e^{j\delta_{i}}.\label{eq:voltages_DQ}
\end{align}
In order to combine the generator equations with the network, we consider
each generator as a current injection source given below
\begin{equation}
i_{gi}=i_{Qi}+ji_{Di}+\frac{1}{r_{ai}}v_{gi}=i_{Qi}+ji_{Di}+\frac{1}{r_{ai}}(v_{Qi}+jv_{Di}).\label{eq:ig}
\end{equation}
Now, by defining the vector $I_{N\times1}$ consisting of the elements
$I_{j}$ where $I_{j}=i_{gi}$ if the $i-$th generator is connected
to the $j-$th node, otherwise $I_{j}=0$ (load buses), and similarly
the vector $V_{N\times1}$ of the bus voltages, then there is 
\begin{equation}
I=YV\label{eq:admittance}
\end{equation}
where $Y$ is the admittance matrix that includes the summation of
the line, generator and load admittance matrices. By combining the
equations \eqref{eq:Te}-\eqref{eq:admittance} and replacing into
the dynamic model of each generator \eqref{eq:delta}-\eqref{eq:Eq},
one can conclude that the dynamic equations of the entire multimachine
power network with $N$ buses and $n$ generators are given in the
generic nonlinear form
\begin{equation}
\dot{x}=f(x,u)\label{eq:nonlinear_model}
\end{equation}
where the state vector is $x=[\delta_{1}\,...\delta_{n}\,\omega_{1}\,...\omega_{n}\,E'_{q1}\,...E'_{qn}]^{T}$
and the control input vector is $u=[T_{m1}\,...T_{mn}\,E_{f1}\,...E_{fn}]^{T}$.
Hence, the control inputs are the mechanical torque and the field
excitation voltage of each generator. Note that when the $j-$th generator
bus is considered as the reference bus, then $\delta_{j}=0$ and the
state vector size is reduced by one.

\subsection{Control task\label{subsec:Control-task}}

The main goal in a multimachine power network is to ensure that the
grid frequency is maintained at $1pu$ and the bus voltages remain
as close to $1pu$ as possible. These are accomplished by suitable
control design at the mechanical torque and field excitation voltage
of each generator respectively. However, since $n$ generators are
available to accomplish the above tasks, it is often required to additionally
share the network load demand (including line losses) in a proportional
manner, i.e., $n_{1}P_{1}\approx...\approx n_{n}P_{n},$ and $m_{1}Q_{1}\approx...\approx m_{n}Q_{n},$
where $P_{i}$, $Q_{i}$ are the real and reactive powers delivered
by the $i-$th generator and $n_{i}$, $m_{i}$ are the gains that
achieve the desired power sharing.

\begin{figure}[t]
\noindent \centering{}\includegraphics[width=0.37\textwidth]{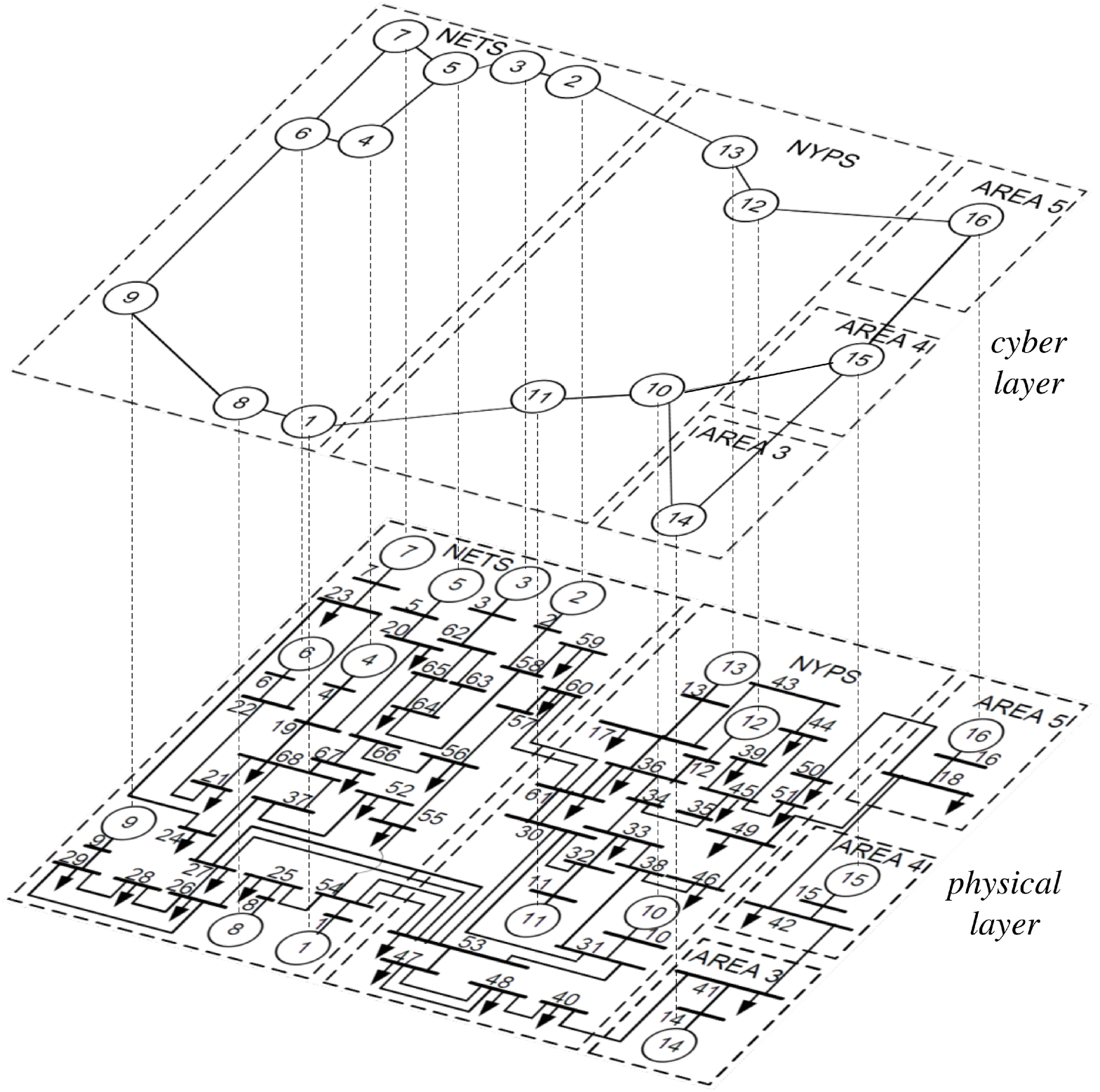}\vspace{-1mm}\caption{Example of the IEEE 68-bus 16-machine power system in a cyber-physical
framework.\label{fig:A-cyber-physical-architecture}}
\vspace{-2mm}
\end{figure}

For the latter, a communication network is established between the
generators in order for the neighbouring units to exchange information
on their real and reactive power. Hence, some necessary properties
for the communication network are denoted. In particular, the communication
network is represented by a connected, undirected and unweighted graph
$G(\mathcal{V},\mathcal{E},\mathcal{A})$, with a set of vertices
$\mathcal{V}=[v_{1}\,\,v_{2}\,\,...\,\,v_{n}]$, connected by a set
of edges $\mathcal{E}\subset\mathcal{V}\times\mathcal{V}$ and induced
by an adjacency matrix $A$. The resulting Laplacian matrix is given
as $L=[A\mathbf{1}_{n}]-A$, where $\mathbf{1}_{n}$ is the vector
all elements equal to 1 and the notation $[w]$ of any vector $w$
represents a diagonal matrix with diagonal entries the elements of
vector $w$. An example of a multimachine power system with a communication
network among the generators is depicted in Fig. \ref{fig:A-cyber-physical-architecture},
where the physical layer includes the IEEE benchmark 68 bus, 16 machine
power system and the cyber layer includes the communication network
among the generator units.

Although the frequency restoration and power sharing can be accomplished
using neighbour-to-neighbour communication with a conventional distributed
integral controller, as widely applied in power networks and microgrids
\cite{7122365,9420296}, the constraints of the input signals (mechanical
torque and field excitation voltage) are often ignored. In particular,
both control input signals should remain within prescribed bounds,
i.e., $T_{mi}\in[T_{mi}^{min},T_{mi}^{max}]$ and $E_{fi}\in[E_{fi}^{min},E_{fi}^{max}]$,
which are given according to the operating limits of the governor
and the operating/thermal limits of the excitation system in order
to maintain synchronism for each generator. Although typical saturation
units can be added to the control inputs, these may result to integrator
windup and instability, while in the case where the system remains
stable, the entire network will no longer operate based on the desired
power sharing operation if the input of one or more generator units
reach the saturation limit. A novel distributed controller that offers
advanced level of resilience in such cases, without requiring a modification
of its dynamics, is presented next.

\noindent 

\section{Resilient distributed control design and analysis\label{sec:Proposed-control-architecture}}

\subsection{Proposed controller\label{subsec:Proposed-controller}}

Consider, initially, the nominal values of the mechanical torque input
$T_{mi}^{n}$ and the field excitation voltage $E_{fi}^{n}$ of each
generator, for which it obviously holds that $T_{mi}^{n}\in(T_{mi}^{min},T_{mi}^{max})$
and $E_{fi}^{n}\in(E_{fi}^{min},E_{fi}^{max})$. In order to accomplish
the desired input constraints without the need of a saturation unit,
the main concept of the original Bounded Integral Control (BIC), proposed
in \cite{Konstantopoulos_Zhong_2016}, is adopted but should be modified
accordingly to achieve the desired control tasks in a distributed
manner, and reduce the number of integral states; thus simplifying
the control implementation. Since the control tasks (frequency restoration
and power sharing) should be accomplished independently of any generator
input reaching its upper or lower limit, the proposed resilient distributed
controller takes the form:
\begin{align}
T_{m} & =T_{m}^{n}+\sigma_{T}\label{eq:sigma_T}\\
E_{f} & =E_{f}^{n}+\sigma_{E}\label{eq:sigma_E}\\
\dot{\sigma}_{T} & =k_{T}\left([g_{T}]\left(\omega_{S}\mathbf{1}_{n}-\omega\right)-k_{P}L_{T}nP\right)-k\sigma_{T}\label{eq:sigma_T_dot}\\
\dot{\sigma}_{E} & =-k_{E}L_{E}mQ-k\sigma_{E}\label{eq:sigma_E_dot}
\end{align}
where $T_{m}=[T_{m1}...T_{mn}]^{T}$, $E_{f}=[E_{f1}...E_{fn}]^{T}$,
$\sigma_{T}=[\sigma_{T1}...\sigma_{Tn}]^{T}$, $\sigma_{E}=[\sigma_{E1}...\sigma_{En}]^{T}$,
$T_{m}^{n}=[T_{m1}^{n}...T_{mn}^{n}]^{T}$, $E_{f}^{n}=[E_{f1}^{n}...E_{fn}^{n}]^{T}$,
$\omega=[\omega_{1}...\omega_{n}]^{T},$$P=[P_{1}...P_{n}]^{T}$,
$Q=[Q_{1}...Q_{n}]^{T}$, $n=diag(n_{1},...n_{n})$, $m=diag(m_{1},...m_{n})$,
scalars $k_{T},k_{P}$, $k_{E}$ are positive constants and $k$ is
selected as an arbitrarily small positive constant. Furthermore,
\begin{align}
g_{T} & =\left[\begin{array}{c}
\left(1-\frac{\sigma_{T1}}{\Delta T_{m1}^{max}}\right)\left(1+\frac{\sigma_{T1}}{\Delta T_{m1}^{min}}\right)\\
\vdots\\
\left(1-\frac{\sigma_{Tn}}{\Delta T_{mn}^{max}}\right)\left(1+\frac{\sigma_{Tn}}{\Delta T_{mn}^{min}}\right)
\end{array}\right],\label{eq:gT}\\
g_{E} & =\left[\begin{array}{c}
\left(1-\frac{\sigma_{E1}}{\Delta E_{f1}^{max}}\right)\left(1+\frac{\sigma_{E1}}{\Delta E_{f1}^{min}}\right)\\
\vdots\\
\left(1-\frac{\sigma_{En}}{\Delta E_{fn}^{max}}\right)\left(1+\frac{\sigma_{En}}{\Delta E_{fn}^{min}}\right)
\end{array}\right],\label{eq:gE}
\end{align}
with $\Delta T_{mi}^{max}=T_{mi}^{max}-T_{mi}^{n}$, $\Delta T_{mi}^{min}=T_{mi}^{n}-T_{mi}^{min}$,
$\Delta E_{fi}^{max}=E_{fi}^{max}-E_{fi}^{n}$, $\Delta E_{fi}^{min}=E_{fi}^{n}-E_{fi}^{min}$,
where $\Delta T_{mi}^{max},\Delta T_{mi}^{min},\Delta E_{fi}^{max},\Delta E_{fi}^{min}>0$,
and 
\begin{align}
L_{T} & =[g_{T}][Ag_{T}]-[g_{T}]A[g_{T}]\label{eq:Laplacian_T}\\
L_{E} & =[g_{E}][Ag_{E}]-[g_{E}]A[g_{E}],\label{eq:Laplacian_E}
\end{align}
where $A$ is the adjacency matrix of the communication network. Note
that in order to have $T_{mi}\in[T_{mi}^{min},T_{mi}^{max}]$ and
$E_{fi}\in[E_{fi}^{min},E_{fi}^{max}]$, from \eqref{eq:sigma_T}-\eqref{eq:sigma_E},
there should be $\sigma_{Ti}\in[-\Delta T_{mi}^{min},\Delta T_{mi}^{max}]$
and $\sigma_{Ei}\in[-\Delta E_{fi}^{min},\Delta E_{fi}^{max}]$. Although
the boundedness of the controller states within these sets will be
analytically shown in the next subsection, it is worth noting that
if $\sigma_{Ti}$ and $\sigma_{Ei}$ are maintained within their given
bounds, then all the elements of vectors $g_{T}$ and $g_{E}$ are
positive. In this case, one can easily observe that matrices $L_{T}$
and $L_{E}$, given in \eqref{eq:Laplacian_T}-\eqref{eq:Laplacian_E},
are symmetric Laplacian matrices.

To further elaborate, let us investigate matrix $L_{T}$. All diagonal
elements are given by the diagonal matrix $[g_{T}][Ag_{T}]$ and are
positive, while the non-diagonal elements are provided from the symmetric
matrix $-[g_{T}]A[g_{T}]$ and are non-positive. In fact, $[g_{T}]A[g_{T}]$
takes the form of a weighted adjacency matrix and has zero diagonal
terms due to the adjacency matrix $A$. Hence, the diagonal elements
of $L_{T}$ are $L_{T(i,i)}=g_{Ti}\stackrel[\underset{(i,j)\in\mathcal{E}}{j=1}]{n}{\sum}g_{Tj}$
and the non-diagonal elements are $L_{T(i,j)}=-g_{Ti}g_{Tj}$, indicating
that $L_{T}$ is a symmetric Laplacian matrix. The same holds for
$L_{E}$.

As for the proposed controller dynamics, the controller for the $i-$th
generator unit takes the form:
\begin{align}
T_{mi} & =T_{mi}^{n}+\sigma_{Ti}\label{eq:sigma_Ti}\\
E_{fi} & =E_{fi}^{n}+\sigma_{Ei}\label{eq:sigma_Ei}\\
\dot{\sigma}_{Ti} & =k_{T}g_{Ti}\left(\!\omega_{S}-\omega_{i}-k_{P}\!\!\stackrel[\underset{(i,j)\in\mathcal{E}}{j=1}]{n}{\sum}(n_{i}P_{i}-n_{j}P_{j})g_{Tj}\right)\nonumber \\
 & -k\sigma_{Ti},\,\,\,\forall i=1,...n\label{eq:sigma_T_doti}\\
\dot{\sigma}_{Ei} & =-k_{E}g_{Ei}\left(\stackrel[\underset{(i,j)\in\mathcal{E}}{j=1}]{n}{\sum}(m_{i}Q_{i}-m_{j}Q_{j})g_{Ej}\right)\nonumber \\
 & -k\sigma_{Ei},\,\,\,\forall i=1,...n\label{sigma_E_doti}
\end{align}
The initial conditions of the controller states should satisfy $\sigma_{Ti}(0)\in(-\Delta T_{mi}^{min},\Delta T_{mi}^{max})$
and $\sigma_{Ei}(0)\in(-\Delta E_{fi}^{min},\Delta E_{fi}^{max})$.

In order to explain how the proposed distributed controller accomplished
the frequency restoration and power sharing when all control inputs
are within their prescribed limits (normal operation), let us consider
the steady-state operation of the system and the control dynamics
where $\omega_{1e}=...=\omega_{ne}$, the real and reactive power
values are $P_{1e},...P_{ne}$, $Q_{1e},...Q_{ne}$ and the controller
states are regulated at the values $\sigma_{T1e},...\sigma_{Tne}$,
$\sigma_{E1e},...\sigma_{Ene}$, where $T_{mie}=T_{mi}^{n}+\sigma_{Tie}\in(T_{mi}^{min},T_{mi}^{max})$
and $E_{fie}=E_{fi}^{n}+\sigma_{Eie}\in(E_{fi}^{min},E_{fi}^{max})$,
for all $i=1,...n$. Then from \eqref{eq:sigma_T_dot} and \eqref{eq:sigma_E_dot}
at the steady state, it leads to
\begin{align*}
[g_{Te}]\left(\omega_{S}\mathbf{1}_{n}-\omega_{e}\right)-k_{P}L_{T}nP & _{e}=\frac{k}{k_{T}}\sigma_{T}\\
-L_{E}mQ_{e} & =\frac{k}{k_{E}}\sigma_{E}
\end{align*}
For an arbitrarily small value of $k$, it yields
\begin{align}
[g_{Te}]\left(\omega_{S}\mathbf{1}_{n}-\omega_{e}\right)-k_{P}L_{T}nP_{e} & \approx0\label{eq:power_sharing_ss}\\
L_{E}mQ_{e} & \approx0.\label{eq:reactive_sharing_ss}
\end{align}
Since $L_{T}$ is a Laplacian matrix and $L_{T}=L_{T}^{T}$, then
$\mathbf{1}_{n}^{T}L_{T}=\mathbf{0}_{n}^{T}$. By left multiplying
\eqref{eq:power_sharing_ss} with $\mathbf{1}_{n}^{T}$ there is
\begin{align}
\mathbf{1}_{n}^{T}[g_{Te}]\left(\omega_{S}\mathbf{1}_{n}-\omega_{e}\right) & \approx0\nonumber \\
g_{Te}^{T}\left(\omega_{S}\mathbf{1}_{n}-\omega_{e}\right) & \approx0.\label{eq:ss_aprox_w-1}
\end{align}
Given that $\omega_{1e}=...=\omega_{ne}$ and all the elements of
$g_{Te}$ are positive, then from \eqref{eq:ss_aprox_w-1} one obtains
that
\begin{equation}
\omega_{1e}=...=\omega_{ne}\approx\omega_{S}\label{eq:frequency_restoration-1}
\end{equation}
leading to the desired frequency restoration. Now from \eqref{eq:ss_aprox_w-1},
after taking into consideration \eqref{eq:frequency_restoration-1},
there is $L_{T}nP_{e}\approx\mathbf{0}_{n}$ which consequently leads
to the desired power sharing $n_{1}P_{1}\approx...\approx n_{n}P_{n}.$
Similarly from \eqref{eq:reactive_sharing_ss} it yields $m_{1}Q_{1}\approx...\approx m_{n}Q_{n}$
which confirms the desired reactive power sharing.

\subsection{Controller boundedness and resilience properties\label{subsec:Controller-boundedness-and}}

To explain how the proposed distributed controller ensures that both
generator inputs $T_{mi}$ and $E_{fi}$, and consequently the controller
states $\sigma_{Ti}$ and $\sigma_{Ei}$, are maintained within the
desired bounds $T_{mi}\in[T_{mi}^{min},T_{mi}^{max}]$ and $E_{fi}\in[E_{fi}^{min},E_{fi}^{max}]$
(equivalently $\sigma_{Ti}\in[-\Delta T_{mi}^{min},\Delta T_{mi}^{max}]$
and $\sigma_{Ei}\in[-\Delta E_{fi}^{min},\Delta E_{fi}^{max}]$),
for every generator unit, the controller dynamics \eqref{eq:sigma_T_doti}
and \eqref{sigma_E_doti} are analysed. It is reminded that initially
there is $\sigma_{Ti}(0)\in(-\Delta T_{mi}^{min},\Delta T_{mi}^{max})$
and $\sigma_{Ei}(0)\in(-\Delta E_{fi}^{min},\Delta E_{fi}^{max})$.

\begin{figure}[t]
\noindent \centering{}\includegraphics[bb=170bp 560bp 370bp 650bp,clip,scale=0.7]{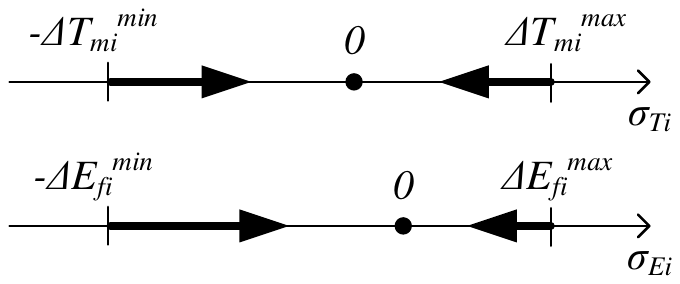}\vspace{-0.2cm}\caption{Illustration of the controller state vector field.\label{fig:Controller-state-vector}}
\vspace{-0.4cm}
\end{figure}

Let us start with the dynamics \eqref{eq:sigma_T_doti} of $\sigma_{Ti}$.
At the upper limit, i.e., when $\sigma_{Ti}=\Delta T_{mi}^{max}$,
the vector field from \eqref{eq:sigma_T_doti} becomes
\begin{equation}
\left.f(\sigma_{Ti})\right|_{\sigma_{Ti}=\Delta T_{mi}^{max}}=-k\Delta T_{mi}^{max}.\label{eq:vector_field_Tmax}
\end{equation}
Expression \eqref{eq:vector_field_Tmax} obviously describes a vector
pointing towards the origin, i.e., inwards towards the closed set
$S_{Ti}=\left\{ \sigma_{Ti}\in R:\,\,-\Delta T_{mi}^{min}\leq\sigma_{Ti}\leq\Delta T_{mi}^{max}\right\} .$
Furthermore, at the lower limit, i.e., when $\sigma_{Ti}=-\Delta T_{mi}^{min}$,
the vector field becomes
\begin{equation}
\left.f(\sigma_{Ti})\right|_{\sigma_{Ti}=-\Delta T_{mi}^{min}}=k\Delta T_{mi}^{min},\label{eq:vector_field_Tmin}
\end{equation}
which also points inwards to $S_{Ti}$. Hence, since initially $\sigma_{Ti}(0)$
belongs in the interior of $S_{Ti}$, then the trajectory $\sigma_{Ti}(t)$
will remain in the interior of $S_{Ti}$ for all future time, i.e.,
$\sigma_{Ti}\in[-\Delta T_{mi}^{min},\Delta T_{mi}^{max}],\,\forall t\geq0.$
A similar vector field analysis for the controller state $\sigma_{Ei}$
can show that $\sigma_{Ei}\in[-\Delta E_{fi}^{min},\Delta E_{fi}^{max}],\,\forall t\geq0$,
resulting in the boundedness of the controller states, and equivalently
the control inputs of the generators, within the predefined ranges.
An illustration of the controller state vector fields at the upper
and lower limits of their bounds is shown in Fig. \ref{fig:Controller-state-vector}.

In order to analyse the resilience property of the proposed distributed
controller, let us consider the case where, at the steady state, due
to a load change, the controller of $l-$th generator (where $l\in\{1,...n\}$)
leads its mechanical torque input to the upper limit, ie. $T_{mle}=T_{ml}^{max}$
or equivalently $\sigma_{Tle}=\Delta T_{ml}^{max}$, while the remaining
equations satisfy the entire load demand and their mechanical torque
inputs remain within the interior of their predefined set. From \eqref{eq:gT},
there is 
\[
g_{Tl}=0\,\,\mbox{and}\,\,g_{Ti}\neq0,\,\forall i=1,...n,\,\mbox{with}\,\,i\neq l.
\]
Then, at the steady state, \eqref{eq:sigma_T_doti} yields for an
arbitrarily small $k$ that
\begin{equation}
\omega_{S}-\omega_{i}-k_{P}\!\!\stackrel[\underset{(i,j)\in\mathcal{E}}{j=1,}j\neq l]{n}{\sum}(n_{i}P_{i}-n_{j}P_{j})g_{Tj}\approx0.\label{eq:ss_without_k}
\end{equation}
This expression is similar to \eqref{eq:power_sharing_ss} where the
$l-$th equation is missing. In addition, the $l-$th generator is
automatically removed from the power sharing since $g_{Tl}=0$ and
the corresponding terms are removed from the summation in \eqref{eq:ss_without_k}.
Hence, the same analysis conducted in Section \ref{subsec:Proposed-controller}
simply proves that the remaining generators, i.e., for $i=1,...n,\,\mbox{with}\,\,i\neq l$,
achieve the desired frequency restoration and power sharing among
them, verifying the resilience property of the proposed distributed
controller. It is underlined that the same would occur if the mechanical
torque input of the $l-$the generator reaches its lower limit, and
consequently the same holds for the field excitation voltage and the
reactive power sharing. The boundedness and the resilient operation
of the entire system is validated in the simulation scenarios that
follow in the next section.

\noindent 
\vspace{-0.4cm}

\section{Simulation results\label{sec:4Simulation-results}}

\begin{figure}[b]
\noindent \centering{}\includegraphics[bb=20bp 20bp 640bp 520bp,clip,scale=0.28]{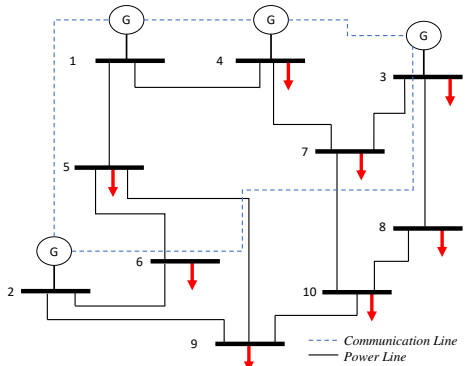}\vspace{-0.2cm}\caption{10-bus, 4-generator power system with the cyber and physical connections.\label{fig:power-system-graph}}
\end{figure}

In order to assess the effectiveness of the proposed controller, a
power system with 4 generators and 10 buses is considered. The system's
physical and communication layers are shown in Fig. \ref{fig:power-system-graph},
while the parameters of the transmission lines are shown in Table
\ref{tab:Power-Lines-Data}. Given the communication graph in Fig.
\ref{fig:power-system-graph} the adjacency matrix of the system is
$A=\left[0,1,0,1;1,0,1,0;0,1,0,1;1,0,1,0\right]$ and the Laplacian
matrices $L_{T}$, $L_{E}$ are calculated using \eqref{eq:Laplacian_T}-\eqref{eq:Laplacian_E}.

\begin{table}[tbh]
\centering{}\includegraphics[bb=320bp 80bp 650bp 460bp,clip,scale=0.4]{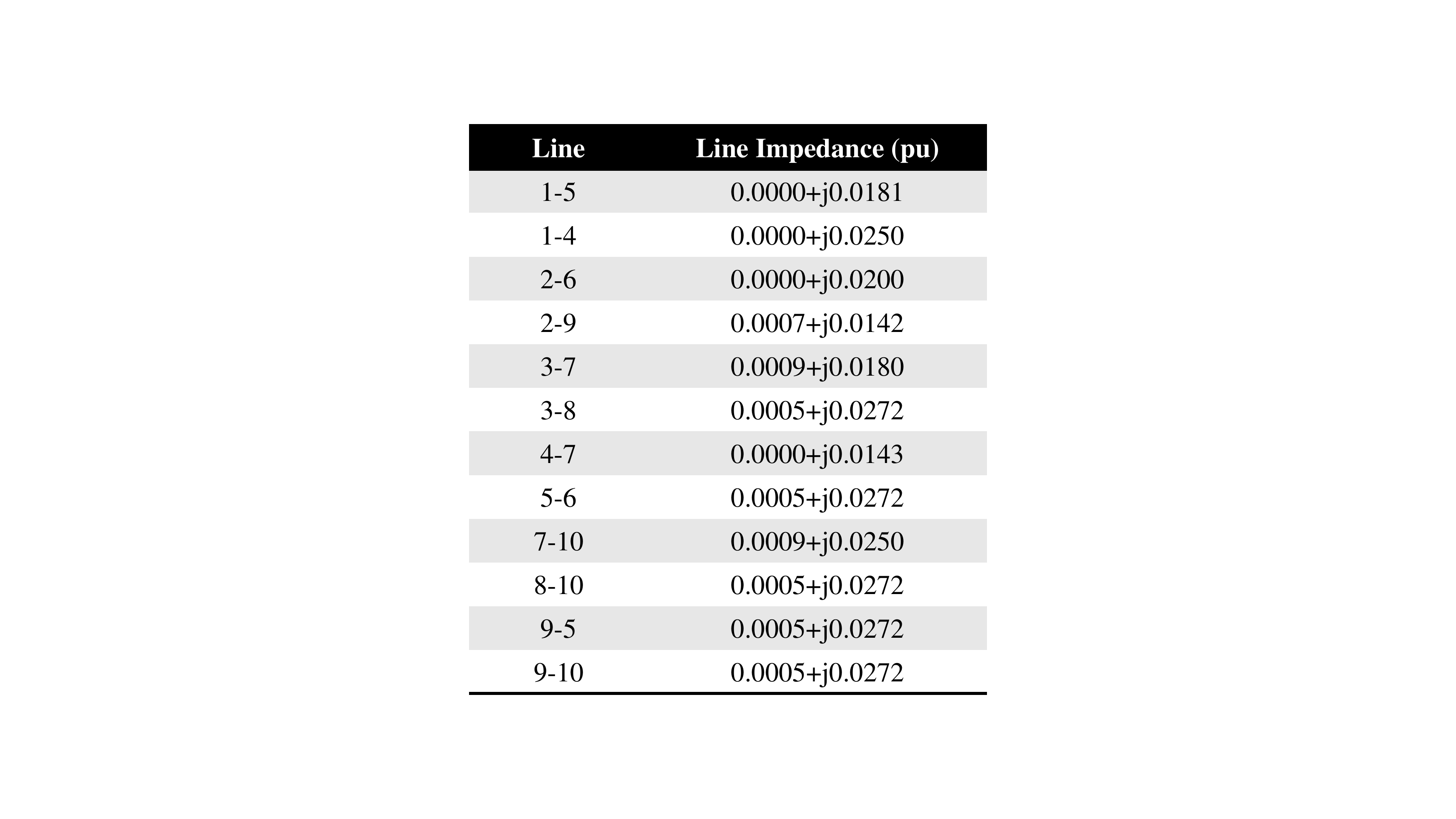}\vspace{-0.2cm}\caption{Power Lines Data \label{tab:Power-Lines-Data}}
\end{table}

For the given multimachine power system, we consider the case where
the torque limits of the generators are set to $\Delta T_{m}^{max}=[2\,\,7\,\,0.5\,\,10]^{T}$
and $\Delta T_{m}^{min}=[2\,\,4.36\,\,4.8\,\,6.1521]^{T}$, while
the field voltage limits are $\Delta E_{f}^{max}=[2\,\,0.2\,\,2\,\,2]^{T}$
and $\Delta E_{f}^{min}=[0.5\,\,0.5\,\,0.5\,\,0.5]^{T}$ (all values
in $pu$). The gains of the controller are set to $k_{T}=160$, $k_{P}=0.0025$,
$k_{E}=0.1$ and $k=10^{-6}$. The rated frequency is $\omega_{s}=1pu$
and the gains for the real and reactive power sharing are $n=diag\left(\frac{1}{2},\frac{1}{4},\frac{1}{6},\frac{1}{8}\right)$
and $m=diag\left(1,1,1,1\right)$, respectively, aiming at the desired
sharing of $P_{1}:P_{2}:P_{3}:P_{4}=1:2:3:4$ and $Q_{1}:Q_{2}:Q_{3}:Q_{4}=1:1:1:1$,
respectively.

At the beginning of the simulation, the power system is at the steady
state with each generator operating at the rated frequency $\omega_{s}$
based on constant inputs $T_{mi}$ and $E_{fi}$ calculated from the
load flow analysis. At time $t=10s$ the proposed distributed controller
is activated. At time $t=300s,$ the load demand is increased, represented
by increasing the load in terms of real power at bus 4 by $5pu$,
which is then disconnected (load shedding) at time $t=600s$.

\begin{figure}[b]
\noindent \centering{}\includegraphics[scale=0.565]{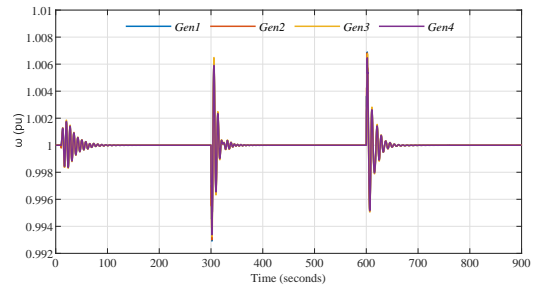}\caption{Generator frequencies.\label{fig:generator-frequencies}}
\end{figure}

\begin{figure}[tbh]
\subfloat[Generator real power output\label{fig:generator-power}]{\noindent \centering{}\includegraphics[scale=0.565]{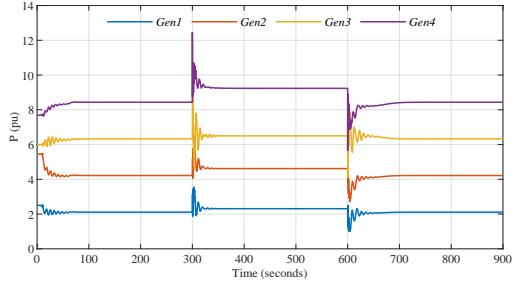}}

\vspace{-0.4cm}\subfloat[Generator reactive power output\label{fig:generator-react-power}]{\noindent \centering{}\includegraphics[scale=0.565]{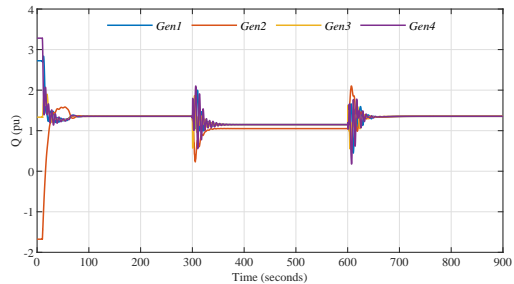}}

\vspace{-0.1cm}\caption{Generator ouputs.\label{fig:generator-ouputs}}
\end{figure}

\begin{figure}[tbh]
\subfloat[Generator mechanical torque\label{fig:generator-torque-input}]{\begin{centering}
\includegraphics[scale=0.565]{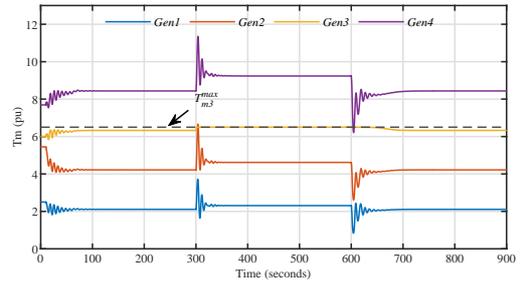}
\par\end{centering}
}

\vspace{-0.4cm}\subfloat[Generator field voltage\label{fig:generator-field-voltage}]{\begin{centering}
\includegraphics[scale=0.565]{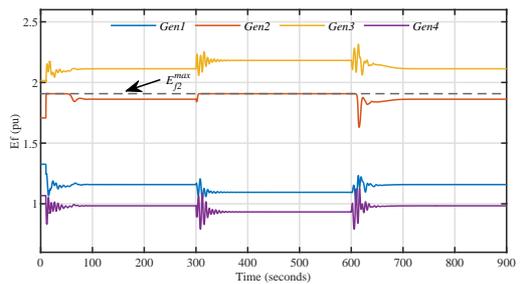}
\par\end{centering}
}\vspace{-0.1cm}\caption{Generator inputs.\label{fig:generator-inputs}}
\end{figure}

As shown in Figs. \ref{fig:generator-frequencies}, \ref{fig:generator-power},
and \ref{fig:generator-react-power}, when the controller is activated,
the real and reactive power sharing is achieved based on the specified
droop gains and the generators operate at rated frequency $\omega_{s}$.
When the load increases, the torque input of generator 3 reaches its
upper bound without exceeding it, as already proven in the theory
and demonstrated in Fig. \ref{fig:generator-torque-input}. Generator
3 is automatically excluded from the power sharing procedure, while
the remaining generators continue sharing real power with $P_{1}:P_{2}:P_{4}=1:2:4$
and restore the frequency back to its rated value ($1pu$), as is
evident in Figs. \ref{fig:generator-frequencies} and \ref{fig:generator-power}.
Similarly, the increase of the load causes generator 2 to reach its
upper field voltage limit (Fig. \ref{fig:generator-field-voltage})
and generator 2 is automatically excluded from the reactive power
sharing procedure, while generators 1, 3 and 4 preserve the reactive
power sharing among them, ie. $Q_{1}:Q_{3}:Q_{4}=1:1:1$ (Fig. \ref{fig:generator-react-power}).
This verifies the resilient operation of the proposed controller.
Finally, as shown in Figs. \ref{fig:generator-power} and \ref{fig:generator-react-power},
when the load decreases back to its initial value, the active and
reactive power sharing between all four generators is restored and
the frequency is also restored to its rated value.

\section{Conclusions\label{sec:5Conclusions}}

A resilient distributed integral controller is introduced in this
paper in order to achieve frequency restoration and power sharing
among multiple generator units and at the same time ensure that all
generator inputs remain within prescribed bounds. Even under scenarios
that force a generator input to operate at its limit, the proposed
controller ensures that the remaining generators share the entire
load proportionally and restore the grid frequency without requiring
any modification of the controller structure or dynamics. The controller's
efficiency and resiliency have been analytically proven and then validated
in a simulated environment of a 10-bus, 4-generator system.

\bibliographystyle{IEEEtran}
\bibliography{datab,George_2,Yiannis}

\end{document}